\documentclass[12pt,unisize]{article}
\begin{document}

\begin{center}
GRAVITATIONAL ULTRARELATIVISTIC INTERACTION\\
OF CLASSICAL PARTICLES IN THE CONTEXT\\
OF UNIFICATION OF INTERACTIONS\\

\bigskip

Roman Plyatsko\\
Pidstryhach Institute for Applied Problems in Mechanics and Mathematics,\\
National Academy of Sciences,
Naukova St. 3-b, Lviv, 79060, Ukraine\\

\medskip

Oleksa Bilaniuk\\
Department of Physics and Astronomy, Swarthmore College,\\
500 College Ave, Swarthmore, Pennsylvania 19081-1397, USA\\

\end{center}

\bigskip

\begin{abstract}
The dependence of the gravitoelectric and gravitomagnetic field components
on the relative velocity of a Schwarzschild source and a local observer
is considered. Three kinds of these components are identified at the
ultrarelativistic velocity, namely, which are proportional to $\gamma^2$,
$\gamma$, and which are independent of $\gamma$ ($\gamma$ is the
relativistic Lorentz factor). The physical situations which evince the
roles of different components are described. Particularly the reaction
of spin on the ultrarelativistic gravitomagnetic field is analysed.
A tendency of gravitational and electromagnetic interactions to approach
in quantitative terms at ultrarelativistic velocities is discussed.
\end{abstract}


\section{Introduction}
The theory of unification of interactions has as its aim to elucidate
the basic properties of the micro-world at high energies. An interesting
question is the
following: is there not a tendency of gravitational and electromagnetic
interactions to approach (at least in quantitative terms) already at a
macro-level in situations where the interacting objects have very high
relative velocities? In other words, are we not getting an indication,
in the framework of general theory of relativity and classical
electromagnetism, of a tendency of diminution of the difference between
these interactions when relative velocities of the interacting classical
particles are ultrarelativistic?

Certainly, investigations of analogies between gravitational and
electromagnetic interactions have their own long history on the level
of Newton's Law of universal gravitation and Coulomb's Law. In the last
decades fundamentally new common traits of these interactions were brought
to light. In particular, terms such as "gravitoelectric field",
"gravitomagnetic field", and "gravitoelectromagnetism" have gained currency
within the general theory of relativity, along with the elaboration of
their context, \cite{1}. Particularly worthy of attention is an
early publication \cite{2}.

Interesting aspects of analogies between gravitation and electromagnetism
are revealed when one looks at the behavior of a classical test particle
with spin in a gravitational field \cite{3}. Examining the
Mathisson-Papapetrou (MP) equations \cite{4} in a Kerr field, \cite{3}
considers gravitational spin-orbit and spin-spin interactions in respective
approximations in power of $1/c$ and compares them with analogous
electromagnetic interactions.

The concept of the electromagnetic field was preceded by concepts of two
independent entities, the electric and the magnetic interaction, which
became unified in Maxwell's theory. On the other hand, in the case of
Einstein's theory of gravitation just the opposite took place: the general
theory of relativity was from the start a theory of a single gravitational
field and only with time did many investigators begin to feel the need to
treat as separate (although closely linked) two of its components:
gravitoelectric and gravitomagnetic. The separation of gravitoelectric and
gravitomagnetic components of the gravitational field in the general theory
of relativity is carried out with the Riemann tensor as the basic
characteristic of the field \cite{2,5}. In the local orthonormal basis,
the gravitoelectric components of the gravitational field $E_{(k)}^{(i)}$ are
determined by the relationship
\begin{equation}\label{1}
E_{(k)}^{(i)}=R^{(i)(4)}_{}{}{}{}{}{}_{(k)(4)},
\end{equation}
where $R^{(i)(4)}_{}{}{}{}{}{}_{(k)(4)}$ denotes local components
of the Riemann tensor. (Here and
in the following the indices of the orthogonal tetrads are placed in round
parentheses; Latin indices run through values 1, 2, 3, while Greek indices
through 1, 2, 3, 4). Correspondingly, for gravitomagnetic components
$B_{(k)}^{(i)}$ we have
\begin{equation}\label{2}
B_{(k)}^{(i)}=-\frac12 R^{(i)(4)}_{}{}{}{}{}{}_{(m)(n)}
\varepsilon^{(m)(n)}_{}{}{}{}{}{}_{(k)},
\end{equation}
where $\varepsilon^{(m)(n)}_{}{}{}{}{}{}_{(k)}$ is the Levi-Civita tensor.

We shall begin our investigation by analyzing relationships (\ref{1})
and (\ref{2}) in the concrete case of a gravitational field created by
a Schwarzschild mass moving relative to an observer with arbitrary
velocity.

   We point out that in interesting paper \cite{6} the similar to a certain
extent problem was considered. Namely, "...it is shown that the gravitational
field of a fast-moving mass bears an increasing resemblance to a plane
gravitational wave, the greater the speed of the mass"  \cite{6}, p. 96.
However, in this paper the gravitational field of a fast-moving Schwarzschild
mass was considered only in the context of investigations of the
gravitational waves in the general theory of relativity. The influence
of components (\ref{1}), (\ref{2}) on the other masses was not under
investigation in \cite{6}.

The gravitational field of a massless particle which moves with the velocity
of light was considered in \cite{7}. It was shown that the gravitational
field of this particle is nonvanishing only on a plane containing the
particle and orthogonal to the direction of motion. The results of \cite{7}
were used in \cite{8} for investigating the ultrarelativistic collision of
two black holes. The detail elucidation of this problem one can find in
\cite{9}. Passing from the infinite Lorentz $\gamma$- factor (the case of
a massless particle with the velocity of light) to the large but finite
$\gamma$ (a massive ultrarelativistic particle) is described taking into
account the small parameter $\gamma^{-1}$ and the corresponding small
corrections to the metric of a massless particle \cite{9}. This approach to
the description of the dependence of the gravitational field of a moving
massive particle on the $\gamma$ differs from the method and results of
\cite{6}. In \cite{6} the Riemann tensor components were calculated for any
value $1< \gamma < \infty$ without the consideration of the case of the
infinite $\gamma$ as the initial approximation.

In what follows, we shall use a system of units where $c=G=1$.

\section{Dependence of the gravitoelectromagnetic field of a moving
Schwarzschield source on the Lorentz $\gamma$-factor}
The results of this Section may be considered as a direct development of
\cite{6}. The main idea of \cite{6} is the comparison of the canonical
forms of the Riemann tensor for a plane gravitational wave and a
fast-moving mass (the $3\times 3$ matrices $P$ and $Q$ in the notation of
\cite{6}). Our purpose is the analysis of the action of different
components (\ref{1}), (\ref{2}), as measured by a fast-moving observer
in the Schwarzschild field, on the test masses.

We shall label the reference frame, which moves with respect to the source
of the Schwarzschild field in an arbitrary direction and with arbitrary
velocity, by a set of corresponding terads $\lambda^\alpha_{(\beta)}$. The
 Schwarzschild metric we consider in standard coordinates $x^1=r, x^2=\theta,
x^3=\varphi, x^4=t$. For expediency and without loss of generality we assume
the directions of the space axes of the ortho-reference to be as follows:
The first axis is perpendicular to the plane determined by the direction of
observer motion and the radial direction to the field source. (Obviously,
in the particular case of radial motion there is freedom of choice). The
second axis coincides with the direction of motion. As a consequence, we
note that the following tetrad components have zero components:
$\lambda^1_{(1)}, \lambda^3_{(1)}, \lambda^2_{(2)}, \lambda^2_{(3)}$. For
evaluation of other components we shall use the general relationship
between terad components and the metric tensor $g^{\alpha\pi}$:
\begin{equation}\label{3}
 \lambda^\alpha_{(\beta)}\lambda^\pi_{(\rho)}\eta^{(\beta)(\rho)}=
 g^{\alpha\pi},
\end{equation}
where $\eta^{(\beta)(\rho)}=diag (-1,-1,-1,1)$ is the Minkowski tensor.

For the Schwarzschild metric, where only the diagonal elements of the
tensor $g^{\alpha\pi}$ are different from zero, the system of the ten
algebraic equations involving tetrad components of (\ref{3}) may be
separated into subsystems of lower dimensionality, permitting to determined
all components that are different from zero:
$$
\lambda^2_{(1)}=\sqrt{-g^{22}},\qquad
\lambda^1_{(2)}=u^1u^4\sqrt{\frac{g_{44}} {u_4u^4-1}},\qquad
\lambda^3_{(2)}=u^3u^4\sqrt{\frac{g_{44}}{u_4u^4-1}},
$$
$$
\lambda^4_{(2)}=\sqrt{\frac{u_4u^4-1}{g_{44}}},\qquad
\lambda^1_{(3)}=u^3\sqrt{\frac{g^{11}g_{33}}{u_4u^4-1}},\qquad
\lambda^3_{(3)}=-u^1\sqrt{\frac{g^{33}g_{11}}{u_4u^4-1}},
$$
\begin{equation}\label{4}
\lambda^1_{(4)}=u^1,\qquad \lambda^3_{(4)}=u^3,\qquad
\lambda^4_{(4)}=u^4,
\end{equation}
where $u^\mu$ is the 4-vector of the observer velocity. (The $\theta$ angle
is measured such that the observer moves in the plane $\theta=\pi/2$, which
entails $u^2=0$. Expressions (\ref{4}) were used in \cite{7} while
dealing with another problem).

According to (\ref{1}) and (\ref{2}), in order to evaluate
$E_{(k)}^{(i)}$ and $B_{(k)}^{(i)}$ it is necessary to have the values of the
local components of the Riemann tensor, which are connected to its
global components by the well known relation
\begin{equation}\label{5}
R_{(\alpha )(\beta )(\gamma )(\delta )}
= \lambda^\mu_{(\alpha )}\lambda^\nu_{(\beta )}
\lambda^\rho_{(\gamma )}\lambda^\sigma_{(\delta )} R_{\mu\nu\rho\sigma}.
\end{equation}
Non-zero components of the Riemann tensor, expressed in standard
Schwa\-rzschild coordinates for $\theta=\pi/2$, are
$$ R_{1212}=R_{1313}=\frac{m}{r-2m},\qquad
R_{2323}=-2mr,
$$
\begin{equation}\label{6}
R_{1414}=\frac{2m}{r^3},\qquad
R_{2424}=R_{3434}=-\frac{m}{r}\left(1-\frac{2m}{r}\right).
\end{equation}
Using (\ref{4})--(\ref{6}), we find the following non-zero
components of the Riemann tensor that are present in (\ref{1}) and
(\ref{2}):
$$
R^{(1)(4)}_{}{}{}{}{}{}_{(1)(4)}=-\frac{m}{r^3}(3u_3u^3-1), \qquad
R^{(2)(4)}_{}{}{}{}{}{}_{(2)(4)}=-\frac{2m}{r^3}\frac{u^1u^1}{g_{44}(u_4u^4-1)}
$$
$$
+\frac{m}{r}\frac{u^3u^3}{u_4u^4-1}, \qquad
R^{(2)(4)}_{}{}{}{}{}{}_{(3)(4)}=R^{(3)(4)}_{}{}{}{}{}{}_{(2)(4)}
=-\frac{3m}{r^2}
\frac{u^1u^3u^4}{u_4u^4-1},
$$
$$
R^{(3)(4)}_{}{}{}{}{}{}_{(3)(4)}=-\frac{m}{r^3}(u_4u^4-1)+
\frac{2m}{r^3}u_4u^4\frac{u_3u^3-u_1u^1}{u_4u^4-1},
$$
$$
R^{(1)(4)}_{}{}{}{}{}{}_{(1)(2)}=\frac{3mu^3u^3u^4}
{r\sqrt{u_4u^4-1}}\left(1-\frac{2m}{r}\right)^{1/2},
$$
$$
R^{(1)(4)}_{}{}{}{}{}{}_{(1)(3)}=-\frac{3mu^1u^3}
{r^2\sqrt{u_4u^4-1}}\left(1-\frac{2m}{r}\right)^{-1/2},
$$
$$
R^{(2)(4)}_{}{}{}{}{}{}_{(2)(3)}=\frac{3mu^1u^3}
{r^2\sqrt{u_4u^4-1}}\left(1-\frac{2m}{r}\right)^{-1/2},
$$
\begin{equation}\label{7}
R^{(3)(4)}_{}{}{}{}{}{}_{(2)(3)}=\frac{3mu^3u^3u^4}
{r\sqrt{u_4u^4-1}}\left(1-\frac{2m}{r}\right)^{1/2}.
\end{equation}
Using (\ref{7}) in (\ref{1}), we obtain the following non-zero
components of the gravitoelectric field:
$$
E^{(1)}_{(1)}=\frac{m}{r^3}(1+3u_\perp^2),\qquad
E^{(2)}_{(2)}=-\frac{2m}{r^3}+\frac{3m}{r^3}\frac{u_\perp^2}{u_4u^4-1},
$$
\begin{equation}\label{8}
E^{(2)}_{(3)}=E^{(3)}_{(2)}=-\frac{3m}{r^3}\frac{u_\parallel u_\perp u^4}
{u_4u^4-1},\qquad E^{(3)}_{(3)}=\frac{m}{r^3}-
\frac{3m}{r^3}\frac{u_\perp^2 u_4 u^4}{u_4u^4-1}.
\end{equation}
where $u_\parallel=u^1$ is the radial component of the 4-velocity,
$u_\perp=ru^3$ is its tangential component. Because of the condition
$u_\mu u^\mu=1$, here we have the following relationship:
\begin{equation}\label{9}
u_4u^4-1=u_\perp^2
+\left(1-\frac{2m}{r}\right)^{-1}u_\parallel^2.
\end{equation}
Similarly, using (\ref{7}) in (\ref{2}), we obtain the non-zero
components of the gravitomagnetic field,
$$
B^{(1)}_{(2)}=B^{(2)}_{(1)}=
\frac{3mu_\parallel u_\perp}
{r^3\sqrt{u_4u^4-1}}\left(1-\frac{2m}{r}\right)^{-1/2},
$$
\begin{equation}\label{10}
B^{(1)}_{(3)}=B^{(3)}_{(1)}=
\frac{3m u_\perp^2 u^4}
{r^3\sqrt{u_4u^4-1}}\left(1-\frac{2m}{r}\right)^{1/2}.
\end{equation}
Let us stress that relationships (\ref{8}) and (\ref{10}) hold true for
any arbitrary velocity of the observer.

We shall begin the examination of the components of (\ref{8}) by simply
noting that they have non-zero values even in the Newtonian limit, when
$|u_\parallel|\ll 1$, $|u_\perp|\ll 1$, $u^4\approx 1$. This had to be
expected, inasmuch as in the Newtonian theory there is correspondence
between the $E^{(i)}_{(k)}$ components and the so-called tidal
matrix $E_{ij}$, where
\begin{equation}\label{11}
E_{ij}=-\frac{\partial}{\partial x_i}\frac{\partial}{\partial x_j}
\varphi (\vec x, t),
\end{equation}
i.e. the second derivatives of the Newtonian potential \cite{8}. The
denomination $E_{ij}$ is not fortuitous in view of the fact that the
components of the tidal acceleration $a_{i_{tidal}}$ in the Newtonian
theory are, see \cite{8},
\begin{equation}\label{12}
a_{i_{tidal}}=E_{ij}r_j.
\end{equation}
In the general theory of relativity the components $E_{(k)}^{(i)}$ are also
linked with the tidal acceleration, more exactly with the equation of
deviation of geodesic lines. Taking (\ref{1}) into account, this may be
written as
\begin{equation}\label{13}
\frac{D^2l^{(i)}}{ds^2}=E^{(i)}_{(k)}l^{(k)},
\end{equation}
where $s$ is the proper time, $l^{(i)}$ is the vector of relative deviation
of two neighboring geodesic lines. It is the equation of deviation of
geodesics that was used in \cite{9}, Sec. 31.2, for the analysis of
tidal forces felt by an observer while falling onto a Schwarzschild black
hole.

According to (\ref{13}) we have
\begin{equation}\label{14}
a^{(i)}_{tidal}=E^{(i)}_{(k)}l^{(k)},
\end{equation}
Let us note that in  \cite{9}, Sec. 31.2,
the components $E_{(k)}^{(i)}$ are
not explicitly mentioned in the analysis of tidal force, but only Riemann
tensor components which, according to (\ref{1}), correspond to
$E_{(k)}^{(i)}$. (At another place in \cite{9}, in Sec. 1.6, there is
mention of the analogy between one part of the Riemann tensor components
and the electric field components, and between the other part and the
magnetic field components, but the relationships (\ref{1}) and (\ref{2})
are not given explicitly). Let us also note that in \cite{9}, Sec. 31.2,
the analysis of tidal forces is limited to the case of radial motion, when
$u_\perp=0$. For such motion, according to (\ref{8}), the components
$E_{(k)}^{(i)}$ assume the following values:
\begin{equation}\label{15}
E^{(1)}_{(1)}=\frac{m}{r^3},\quad
E^{(2)}_{(2)}=-\frac{2m}{r^3},\quad
E^{(2)}_{(3)}=E^{(3)}_{(2)}=0,\quad
E^{(3)}_{(3)}=\frac{m}{r^3},
\end{equation}
that is, they appear completely independent of $u_\parallel$. In view of
(\ref{14}), in such a case the tidal acceleration also does not depend
on the velocity of radial motion. The fact that in a radial fall of the
observer the tidal forces felt by him are independent of this velocity is,
in essence, noted in \cite{9}, Sec. 31.2, (while noting at the same
time the analogy with electromagnetism). As a consequence, it is stated
in \cite{9} that for a radial falling observer the tidal forces
increase sharply only at $r\to 0$. The question which remaind unanswed in
\cite{9} was the following: what will change if the fall is non-radial?
In view of expressions (\ref{8}), this question can be readily answered.
According to (\ref{8}), the expressions giving the components of the
gravitoelectric field in the case of non-radial motion differ significantly
from expressions (\ref{15}) only when the velocity becomes ultrarelativistic.
Indeed, inasmuch as $u_\parallel$, $u_\perp$, $u^4$ are proportional to the
Lorentz relativistic $\gamma$-factor, (\ref{8}) gives us, for
$|u_\perp|\gg 1$, $|u_\parallel|\gg 1$, $u^4\gg 1$
\begin{equation}\label{16}
E^{(1)}_{(1)}\approx \frac{3m}{r^3}\gamma^2,\quad
E^{(2)}_{(2)}\approx \frac{3m}{r^3},\quad
E^{(2)}_{(3)}=E^{(3)}_{(2)}\sim \frac{3m}{r^3}\gamma,\quad
E^{(3)}_{(3)}\sim \frac{3m}{r^3}\gamma^2.
\end{equation}
Comparing (\ref{16}) with (\ref{15}), we see that while the
components in (\ref{15}) assume arbitrarily large values only
when $r\to 0$, the components $E_{(1)}^{(1)}$, $E_{(3)}^{(2)}=E_{(2)}^{(3)}$,
$E_{(3)}^{(3)}$ of (\ref{16}) become arbitrarily large already at finite
values of $r$, provided $\gamma\to \infty$. (Here we leave aside the
question of how to impart to an observer, in practice, a velocity
corresponding to large values of $\gamma$). Thus, according to (\ref{14}),
an observer in a Schwarzschild field runs the risk of being torn apart by
tidal forces not only at $r\to 0$, i.e. under the surface of the horizon
(which is described in \cite{9}, Sec. 31.2), but even at large values
of $r$ if his velocity becomes ultrarelativistic.

Obviously, the expressions for the components of the gravitoelectric field
(\ref{8}) and (\ref{16}) are independently valid, without having to be
linked with equations of deviation of geodesic lines and tidal forces. The
significance of (\ref{8}) and (\ref{16}) resides primarily in
characterizing the gravitational field created by a moving Schwarzschild
source.

Let us now look at the components of the gravitomagnetic field,
(\ref{10}). It is easy to see that components (\ref{10}) are different
from zero only when $u_\perp\ne 0$, that is only when the observer is
moving non-radially. (As is well known, a similar situation arises in
electrodynamics for components of the vector of magnetic field intensity
of a moving electric charge). Quite generally, the values of components
(\ref{10}) depend signoficantly on observer motion. In the low relativistic
region, with $|u_\perp|\ll 1$, $|u_\parallel|\ll 1$, $u^4\approx 1$, the
common multiplier $m/r^3$ of components is further multiplied by
corresponding small factors. Whereas in the ultrarelativistic region, where
$|u_\perp|\gg 1$, $|u_\parallel|\gg 1$, $u^4\gg 1$, this multiplier is
further multiplied by large factors, because in this case, according to
(\ref{10}), we have:
\begin{equation}\label{17}
B^{(1)}_{(2)}=B^{(2)}_{(1)} \sim \frac{3m}{r^3}\gamma,\quad
B^{(1)}_{(3)}=B^{(3)}_{(1)} \sim \frac{3m}{r^3}\gamma^2.
\end{equation}
As we have seen, in the Newtonian limit the components of the
gravitomagnetic field (\ref{10}) have zero values, in contrast to the
gravitoelectric field. Moreover, at low relativistic velocities the absolute
values of components (\ref{10}) are considerably smaller than the
components (\ref{8}). Yet at ultrarelativistic velocities , the largest
components $B_{(k)}^{(i)}$ from (\ref{17})
and $E_{(k)}^{(i)}$ from (\ref{16}) are
of the same order of magnitude, determined by the factor $3m\gamma^2/r^3$.
(We point out that this factor is present in the expression for the
amplitude of the gravitational wave from \cite{6}).
Consequently, we can conclude that the two aspects of the single
gravitational interaction, the gravitoelectric and the gravitomagnetic, show,
in the ultrarelativistic range, a tendency of qualitatively drawing
together, even though at low velocities they differ substantially.

Befor comparing the gravitational interaction with the electromagnetic in
the ultrarelativistic range, we shall examine an important physical
situation which evinces the role of the gravitomagnetic interaction.

\section{A classical particle with spin in an ultrarelativistic
gravitational field}
Relationships (\ref{13}), (\ref{14}) describe a simple experiment in
which an observer moving in a Schwarzschild field can determine the values
of components of the gravitoelectric field in his own frame of reference. The
question arises which experiment would permit this observer to determine
the components of the gravitomagnetic field. As is shown in \cite{7},
such an experiment can be carried out by observing the behavior of a test
particle with spin. Indeed, according to (\ref{9}) in \cite{7}, the
local components of the 3-acceleration $a_{(i)}$ with which the test
particle with spin deviates from free geodesic fall in an arbitrary
gravitational field is given by the relationship
\begin{equation}\label{18}
a_{(i)}=-\frac{s_{(1)}}{M}R_{(i)(4)(2)(3)},
\end{equation}
where $M$ is the mass of a test particle.
(Here the space axes of the reference frame are chosen such that the spin
points along the first axis, so that the spin components are
$s_{(2)}=s_{(3)}=0$. Expression (\ref{18}) is a direct cosequence of the
MP equations). Taking (\ref{2}) into consideration, it is not difficult to
see that the right side of (\ref{18}) contains the components of the
gravitomagnetic field.

For our concrete case of observer motion in a Schwarzschild field,
characterized by the set of tetrads (\ref{4}), we have
\begin{equation}\label{19}
a_{(i)}=\frac{s_{(1)}}{M}B^{(1)}_{(i)}.
\end{equation}
This result is obtained from (\ref{18}) taking into account the
appropriate components of the curvature tensor from (\ref{7}) and
expressions (\ref{10}). The non-zero values of $B_{(k)}^{(i)}$ in (\ref{18})
come from (\ref{10}). Even though the right sides of relationships
(\ref{19}) and (\ref{14}) have a similar appearance, what is essential is
that they contain different components of the gravitational field: in
(\ref{19}) gravitomagnetic and in (\ref{14}) gravitoelectric. (Let us
stress that we are dealing with one and the same reference frame, one
connected with an observer moving in a Schwarzschild field). Correspondingly,
the nature of forces that cause accelerations (\ref{14}) and (\ref{19}) is
different: in case of (\ref{14}) these are tidal forces, and in (\ref{19})
it is the spin-orbit force. (Detailed discussion of this question may be
found in \cite{7}). Referring to (\ref{10}) we obtain the magnitude
of the 3-acceleration $|\vec a|$ with components (\ref{19}):
\begin{equation}\label{20}
|\vec a|=\frac{3m}{r^3}\frac{|s_{(1)}u_\perp |}{M}
\sqrt{1+u_\perp^2}.
\end{equation}
According to (\ref{17}), the acceleration components (\ref{19}) depend,
in the case of ultrarelativistic non-radial motion, on the Lorentz
$\gamma$-factor such that $a_{(2)}\sim  \gamma$, $a_{(3)}\sim  \gamma^2$. The
component $a_{(1)}$ remains equal to zero at any velocity in the case at
hand, where the spin is directed along the first spacial vector of the
reference frame, which direction is perpendicular to the plane determined by
the direction of observer motion and the radial direction. This is so because
in this case the corresponding component of the gravitational field is zero
also. Expression (\ref{20}) also shows that $|\vec a|\sim \gamma^2$.

Thus, the fact that at ultrarelativistic velocity the largest component of
the gravitomagnetic field (\ref{17}) is proportional to $\gamma^2$ entails,
on account of (\ref{19}), that the spin-orbit acceleration is also
proportional to $\gamma^2$.

Both, the MP equations and relationships (\ref{18})--(\ref{20}) which
follow from them, are rigorously valid for the model of a point test particle
with spin, with tidal forces not coming into play. Certainly, for any real
macroscopic test particle with rotational motion, tidal and spin-orbit
forces become important.

Together, relationships (\ref{14}) and (\ref{19}) permit to evaluate
these forces. The most significant conclusion of these evaluations lies in
the result that for ultrarelativistic non-radial motions the values of
these forces in the proper frame of the particle are both proportional to
$\gamma^2$.

\section{Comparison of gravitational and electromagnetic interactions of
two particles moving with an ultrarelativistic relative velocity}
We shall examine two situations where two particles are mutually
interacting. In the first case, we consider two electrically neutral
particles with the mass of one being considerably grater than the mass of the
other. The particle with smaller mass is endowed with classical spin
(internal angular momentum). Thus, we can consider this particle to be the
test particle with spin, moving in the gravitational field of the more
massive particle, which in its own frame is described by a Schwarzschild
metric.

In the second case, the particles carry electric charge, with one charge
being considerably larger than the other. Moreover, the particle with the
smaller charge has a magnetic moment arising from its internal rotation.
The masses of these particles are such that at low relative velocities, when
the Coulomb Law and Newton's Law of gravitational attraction hold true, the
force of the electric interaction is very much larger than the gravitational
attraction. Again, the particle with the smaller charge and the magnetic
moment may be regarded as the test particle. Thus, we may consider that the
first pair of particles interacts only gravitationally, and the second pair
only electrically.

Let us inquire, in the two cases, how the forces resulting from the
gravitational and the electromagnetic interactions, respectively, depend on
the magnitude of the relative velocities of the particles. We shall assume
that the particles are sufficiently distant one from the other to be able
to neglect the respective gravitational and electromagnetic radiation. In
accordance with the analysis carried out in the preceding sections, in the
first case the force is due to spin and it increases with increasing
relative velocity proportionally to $\gamma^2$, as long as the test mass
is not moving radially, i.e. it is not moving along the line joining the two
masses.

At the same time, in the second case, classical electrodynamics tells us that
the force acting on the test particle with the magnetic field is proportional
to $\gamma$. This means that, no matter how small the gravitational
interaction may be in comparison with the electromagnetic interaction in the
subrelativistic range of velocities, in passing into the ultrarelativistic
range the ratio of the respective forces could, in principle, change to an
extent of both forces becoming of the same order of magnitude, provided
$\gamma$ becomes large enough.

Similar conclusions may be drawn in a third situation, where a model proton
interacts with a model electron. (Here we consider two classical particles
with masses and charges of a proton and an electron, respectively. In this
case, for the description of the gravitational field of the proton as the
more massive particle, ane has to refer to the Reissner-Nordstrom metric,
rather the Schwarzschild metric, but this does not change the conclusion in
principle).

In the Introduction we asked the question, in the framework of general
relativity and classical electrodynamics, if there is a tendency of
gravitational and electromagnetic interactions to approach quantitatively
with increasing relative velocity of interacting particles. We have shown
that this question may be answered in the affirmative.

\section{A case of the ultrarelativistic motion of a classical
spinning particle in a Schwarzschild field and the corresponding solution
of the Dirac equation}
The physical measurements with the ultrarelativistic macroscopic masses are,
at least at present, unattainable. Nonetheless, the results from Sections
3 and 4 are not merely academic. If only because the known fact that the
general covariant Dirac equation passes, in a quasiclassical approximation,
into the MP equations. A concrete problem that should be tackled, is
obtaining solutions of the general covariant Dirac equation in a
Schwarzschild field corresponding to ultrarelativistic electrons.
   It is not difficult to check that the Mathisson-Papapetrou equations
in a Schwarzschild field have a strict partial solution describing the
circular motion of a spinning test particle around the field source on
the orbit with $r=3m$. The relationship between the components of
the particle's 4-velocity $u_\perp\equiv r\dot\varphi$ and the 3-vector
spin component $S_2\equiv S_\theta$ is
\begin{equation}\label{21}
u_\perp = -\frac{3mM}{S_\theta}
\end{equation}
(as above, here we use the standard Schwarzschild coordinates; spin is
perpendicular to the plane of motion $\theta=\pi/2$, therefore $S_1=0$,
$S_3=0$). It is necessary to take into account the condition for a spinning
test particle $|S_0|/Mr\ll 1$ where $|S_0|$ is the value of the spin of
a test particle as measured by the comoving observer \cite{3} (in our case
there is the relation
$|S_\theta|=ru_4 |S_0|$). Therefore, for the value $|u_\perp|$ from (\ref{21})
we have $|u_\perp|\gg 1$, i.e. for the motion on the circular orbit with
$r=3m$
a particle must possess the ultrarelativistic velocity, the higher the spin
is smaller. Formally, at $S_\theta=0$ the value $|u_\perp|$ in (\ref{21}) must
be infinitely large, that is the particle must move with the speed of light.
This fact corresponds to the known result following from the geodesic
equations in a Schwarzschild field: the circular nonisotropic geodesic
orbits exist only at $r>3m$ and, formally, for the motion on the orbit
$r=3m$ a test particle without spin must possess the speed of light. In
practice it means that only the beam of light can move on the orbit with
$r=3m$.

So, according to the MP Eqs. the spin of a test particle allows its
ultrarelativistic motion on the circular orbit $r=3m$. The calculations
of the gravitational spin-orbit acceleration on the orbit $r=3m$ according
to (\ref{20}), (\ref{21}) give
\begin{equation}\label{22}
|\vec a|= \frac{\sqrt{3}}{9m}.
\end{equation}
For the quantitatively comparison, we point out that value
(\ref{22}) is close to the Newtonian value of
the free fall acceleration for the mass $m$ at the distance $r=3m$ (in the
used system of unites this Newtonian acceleration is equal to $1/9m$).

It is interesting that the orbit $r=3m$ is a common solution of the
MP Eqs. at the two known variants of the auxiliary conditions for these
Eqs., namely, the condition of Pirani and Tulczyjew-Dixon \cite{3}. Generally
the solutions of the MP Eqs. at the different condition do not coincide.

It is clear that the solution of the MP Eqs. describing the orbit
$r=3m$ in a Schwarzschild field is interesting mainly in the theoretical
sense because in practice one cannot deal with a macroscopic particle
moving with the ultrarelativistic velocity relatively the field source.
There is much more perspective situation with the high-energy elementary
particles, e.g. electrons or protons. In this connection the question
arises: does the Dirac equation in a Schwarzschild field have a solution
which corresponds, in the certain meaning, to the considered solution of
the MP Eqs. with $r=3m$?
For answer this question let us analyse the components of the 4-spinor
$\Psi_\mu$ which by the known procedure of separation of the variables in the
Dirac equation in a Schwarzschild field (see, e.g., \cite{10}, Ch. 10) take
the form
$$ \Psi_1=
\frac{1}{r\sqrt{2}}R_{-1/2}(r)S_{-1/2}(\theta)exp[i(\sigma t+
m^\prime\varphi)],
$$
$$
\Psi_2= R_{+1/2}(r)S_{+1/2}(\theta)exp[i(\sigma t+
m^\prime\varphi)],
$$
$$
\Psi_3= -R_{+1/2}(r)S_{-1/2}(\theta)exp[i(\sigma t+
m^\prime\varphi)],
$$
\begin{equation}\label{23}
\Psi_4= -\frac{1}{r\sqrt{2}}R_{-1/2}(r)S_{+1/2}(\theta)exp[i(\sigma t+
m^\prime\varphi)].
\end{equation}
For the radial functions $R_{+1/2}(r)$, $R_{-1/2}(r)$ we have the expressions
$$
R_{+1/2}(r)=\frac{1}{\sqrt{r^2-2mr}}\psi_{+1/2}(r)exp\left(-\frac{i}{2}\arctan
\frac{Mr}{\lambda}\right),
$$
\begin{equation}\label{24}
R_{-1/2}(r)=\psi_{-1/2}(r)exp\left(+\frac{i}{2}\arctan
\frac{Mr}{\lambda}\right),
\end{equation}
where $\lambda$ is the parameter of separation of the variables depended
on the orbital moment, $\sigma$ is the value of energy, and the functions
$\psi_{+1/2}(r)$, $\psi_{-1/2}(r)$ can be find from the two differential
equations written in \cite{10}. We shall consider these Eqs. for the
case $mM\gg 1$, that is when the Schwarzschild source is, e.g., an ordinary
(not microscopic) black hole. When perfoming concrete calculations one
can take into account different values of $\sigma$, $\lambda$ and investigate
the corresponding quantum states. Here we consider the case when $\sigma$
and $\lambda$ are equal to the values of the energy and moment of the
classical electron following from the MP Eqs. for the above considered
circular orbit with $r=3m$. Then for the functions $\psi_{+1/2}$,
$\psi_{-1/2}$ we obtain the equations
$$
\frac{d\psi_{+1/2}}{dx}-iA\left(1-\frac{2}{x}\right)^{-1}\psi_{+1/2}+
3^{3/2}\frac{A}{x}\left(1-\frac{2}{x}\right)^{-1/2}\psi_{-1/2}=0,
$$
\begin{equation}\label{25}
\frac{d\psi_{-1/2}}{dx}+iA\left(1-\frac{2}{x}\right)^{-1}\psi_{-1/2}+
3^{3/2}\frac{A}{x}\left(1-\frac{2}{x}\right)^{-1/2}\psi_{+1/2}=0,
\end{equation}
where $x\equiv r/m$, $A\equiv 2m^3M^3/\sqrt{3}$ (Eqs. (\ref{25})
are written for the values of $x$ which are not in the small
neighborhood of $x=2$). The analysis of the solutions of
(\ref{25}) shows that the property $|\psi_{+1/2}|=|\psi_{-1/2}|$
takes place and the maximum value of $|\psi_{\pm 1/2}|$ is
achieved at $x=3$. We stress that just the values $|\psi_{\pm
1/2}|^2$ together with (\ref{23}), (\ref{24}) determine the
probability to find an electron in the certain space region
because for the components of the Dirac carrent $J^\mu$ the
relationship takes place \cite{10}:
$$
J^\mu=\sqrt{2}\left[l^\mu (|\Psi_1|^2+|\Psi_4|^2)+
n^\mu (|\Psi_2|^2+|\Psi_3|^2)\right.
$$
\begin{equation}\label{26}
\left.- m^\mu (\Psi_1\Psi_2^* - \Psi_3\Psi_4^*)-
m^{*\mu}(\Psi_1^* \Psi_2 - \Psi_3^* \Psi_4)\right],
\end{equation}
where $l^\mu, n^\mu, m^\mu, m^{*\mu}$ are the known isotropic
vectors in the Newman-Penrose formalism. Taking into account
(\ref{23}), (\ref{26}) and the relation
$|\psi_{+1/2}|=|\psi_{-1/2}|$ it is easy to find that
$J^\varphi\ne 0$, $J^t\ne 0$, whereas $J^r = 0$, $J^\theta = 0$.
It follows that the current circulates exactly on the circle and
the maximum values of $|J^\varphi|$, $|J^t|$ are achieved at
$r=3m$. The width of the peak of the curve $|\psi_{\pm 1/2}|^2$
decreases when $A$ grows, and at $A\to \infty$ we have the
classical circular orbit with $r=3m$.

So, the considered solution of the Dirac equation in a Schwarzschild
field describes the quantum state corresponding to the classical orbit with
$r=3m$. We point out that the parameters $\sigma$ and $\lambda$ of this
state are equal to the energy and moment of the classical electron on this
orbit. As we stress above the circular orbit
with $r=3m$ is an example when the
gravitational ultrarelativistic spin-orbit acceleration becomes significant.
Further on it is interesting to investigate other solutions of the Dirac
equation which can show the role of the ultrarelativistic gravitation in the
astrophysical processes.

\section{Conclusions}
Relationships (\ref{16}), (\ref{17}), and the conclusions drawn from them
in concrete physical situations described by expressions (\ref{13}),
(\ref{19}), (\ref{20}), point unequivocally to the need for investigators
to direct more attention to the gravitational interaction at
ultrarelativistic relative velocities. In
view of the correspondence principle, there are reasons to infer that some
important relationships of gravitational interaction of classical
(non-quantum) objects will, to a certain extent, hold for particles of the
micro-world, where ultrarelativistic relative velocities and high energies
are ubiquitous.

An important question asks whether the analyses, carried out
above, might not be helpful in delving into the specifics of inclusion
of the gravitational interaction into the scheme of unification of
interactions. We think they might be. If only because a purely classical
examination affords a deeper insight into the gravitational interaction in
the micro-world at high energies.

Not less important is the need to elucidate how the entity which in classical
terms is denoted as "gravitational ultrarelativistic spin-orbit interaction"
should be expressed in the scheme of second quantization.

\end{document}